# The role of collider bias in understanding statistics on racially biased policing

16 July 2020

Norman Fenton[1,2], Martin Neil[1,2], Steven Frazier[3]

[1] Risk and Information Management, Queen Mary University of London, UK
[2] Agena Ltd, Cambridge, UK
[3] Koch Engineered Solutions, USA

**Abstract**

Contradictory conclusions have been made about whether unarmed blacks are more likely to be shot by police than unarmed whites using the same data. The problem is that, by relying only on data of 'police encounters', there is the possibility that genuine bias can be hidden. We provide a causal Bayesian network model to explain this bias – which is called collider bias or Berkson's paradox – and show how the different conclusions arise from the same model and data. We also show that causal Bayesian networks provide the ideal formalism for considering alternative hypotheses and explanations of bias.

## 1. Introduction

Even before the recent George Floyd case, there has been much debate about the extent to which claims of systemic racism are supported by statistical evidence. For example (Ross 2015) claims that unarmed blacks are 3.5 times more likely to be shot by police than unarmed whites when adjusting for relative differences in population size. However, (Fryer 2016) - formally published later as (Fryer 2019) - found that there was no such racial disparity when the data were conditioned on people being stopped by police, and there was a similar conclusion in (Patty and Hanson 2020) that was produced in direct response to public concerns about the Floyd case.

In response to Fryer, (Ross, Winterhalder, and McElreath 2018) argued that Fryer's analysis was compromised because it was essentially an example of Simpson's paradox (Simpson 1951; Bickel, Hammel, and O'Connell 1975; Fenton, Neil, and Constantinou 2019) whereby conclusions based on pooled statistics are reversed when drilling down into relevant subcategories.

A new paper (Knox, Lowe, and Mummolo 2020) explains why Simpson's paradox is not the only statistical explanation for the apparently contradictory conclusions of Ross and Fryer. They identify why relying only on available administrative data (of police encounters) can lead to a misrepresentation of the extent of racially biased policing. Simply put, the conclusions of (Fryer 2019; Patty and Hanson 2020) may be misleading because the data on which their analyses rely (namely police encounters) is 'censored' in a way that hides racial bias. Although they do not identify it by name, this statistical problem is normally referred to as collider bias or Berkson's paradox (Pearl and Mackenzie 2018; Fenton 2020).

In our view the key contribution of (Knox, Lowe, and Mummolo 2020) is that it provides an explicit causal graphical model - in the sense of (Pearl and Mackenzie 2018) - to explain the statistical paradox. However, their subsequent mathematical explanation is difficult for lay people to follow, and therefore the lay explanation provided by (Bronner 2020) – is welcome. But, in turn, the lay explanation is somewhat vague and fails to make clear the causal explanation for the paradox.

We believe that a Bayesian Network (BN) model and inference (Fenton and Neil 2018; Pearl 2000) provides a relatively simple explanation of the paradox (using the same example data as Bronner), while also highlighting the full causal structure of the problem. We will also show that BNs provide the ideal formalism for considering alternative hypotheses and explanations of potential bias.

## 2. The BN model that explains the problem

The causal BN model explaining the collider bias is shown in Figure 1. A BN is a graphical model, consisting of nodes and arcs where the nodes represent variables which may or may not be directly observable and where there is an arc between two nodes if the corresponding variables are causally or statistically linked. The strength of the link, as well as the uncertainty associated with these, is captured as conditional probabilities.

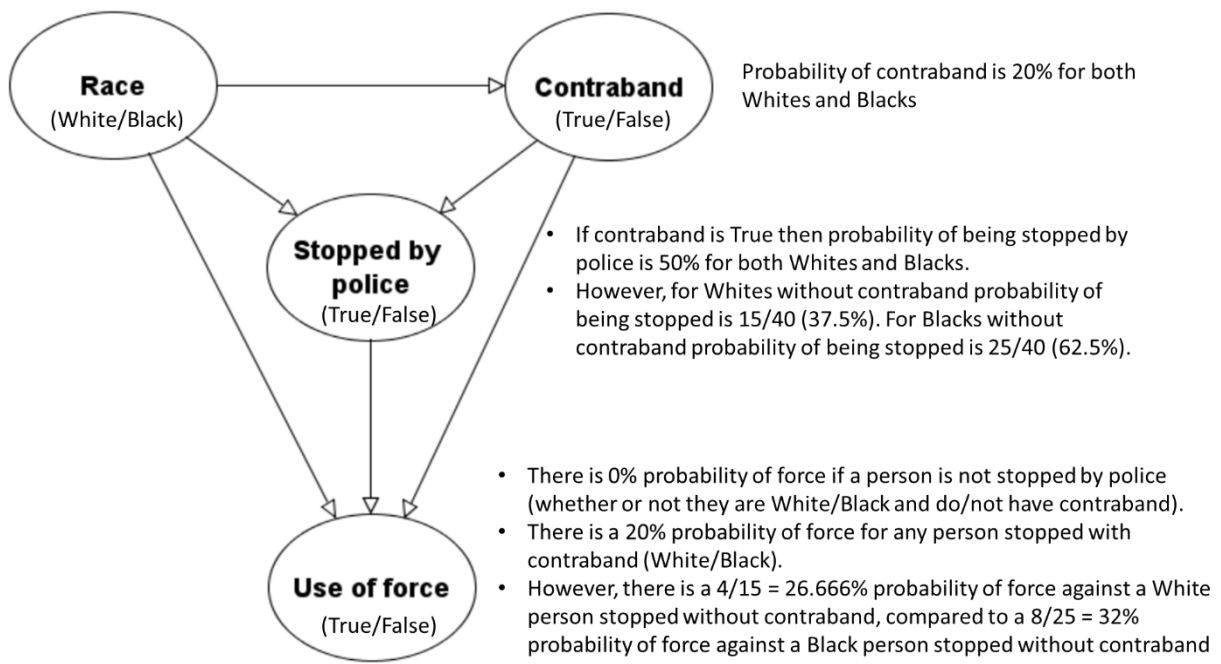

*Figure 1 BN structure with information required for the conditional probability tables*

The notion of 'contraband' (whether or not a person is carrying contraband) was used by Bronner as a concrete example of what (Knox, Lowe, and Mummolo 2020) referred to as 'suspicion'. This is simply the potential trigger event, i.e. 'reason' for a person being stopped by police. So, although Bronner did not consider a causal model, the causal structure of Figure 1 is exactly the same as that assumed by (Knox, Lowe, and Mummolo 2020), but with 'contraband' replacing 'suspicion'.

Bronner uses the hypothetical example of 100 people (50 white, 50 black) which we can think of as population percentages. She assumes:

- An equal number (10) of whites and blacks (i.e. 20% in each case) have contraband. This provides the necessary information for the conditional probability table for the node "Contraband" which is conditioned on the node "Race".
- For anybody with contraband there is a 50% probability of being stopped by police (so it is 50% for both white and blacks). However, blacks without contraband are more likely to be stopped (25/40, i.e. 62.5%) than white without contraband (15/40, i.e. 37.5%). This provides the necessary information for the conditional probability table for the node "Stopped by police" which is conditioned on the nodes "Race" and "Contraband".
- Force is used against 1 in 5 (20%) of people stopped with contraband, irrespective if they are white or black. However, force is more likely to be used against blacks stopped without contraband (8/25 = 32%) than against whites stopped without contraband (4/15 = 26.666%). This provides the necessary information for the conditional probability table for the node "Use of Force" which is conditioned on the nodes "Race", "Contraband" and "Stopped by police".

With these chosen values there is clear bias against blacks: not only are blacks who are not carrying contraband more likely to be stopped than whites not carrying contraband, but they are also more likely to be subject to force. The BN model not only confirms the bias but, more importantly, shows that, *when restricted to data on people stopped, there appears to be no bias*.

We will show these two scenarios in turn.

First note that when we enter the values into the probability tables of the model[1] and run the model we get the prior marginal probabilities shown in Figure 2.

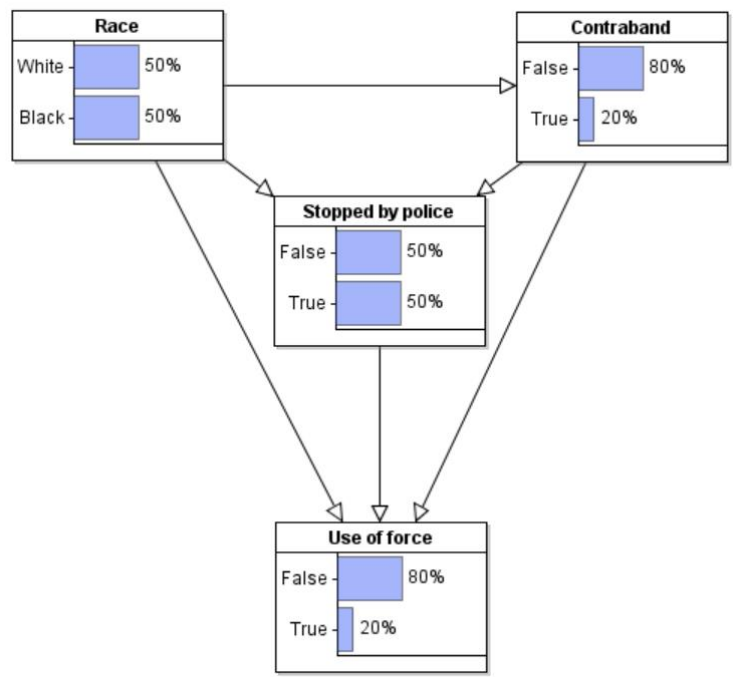

*Figure 2 Prior marginal probabilities*

The fact that there is genuine bias is shown by running the model for whites and blacks respectively as shown in *Figure 3* (when observations are entered into the model for specific variables, all of the probabilities for the, as yet unobserved variables are updated using an AI algorithm called Bayesian inference).

---

[1] We are using the AgenaRisk BN software (Agena Ltd 2020). A free trial version can be downloaded from www.agenarisk.com and the model is available here:
http://www.eecs.qmul.ac.uk/~norman/Models/Race_and_police_encounters_for_web.cmpx

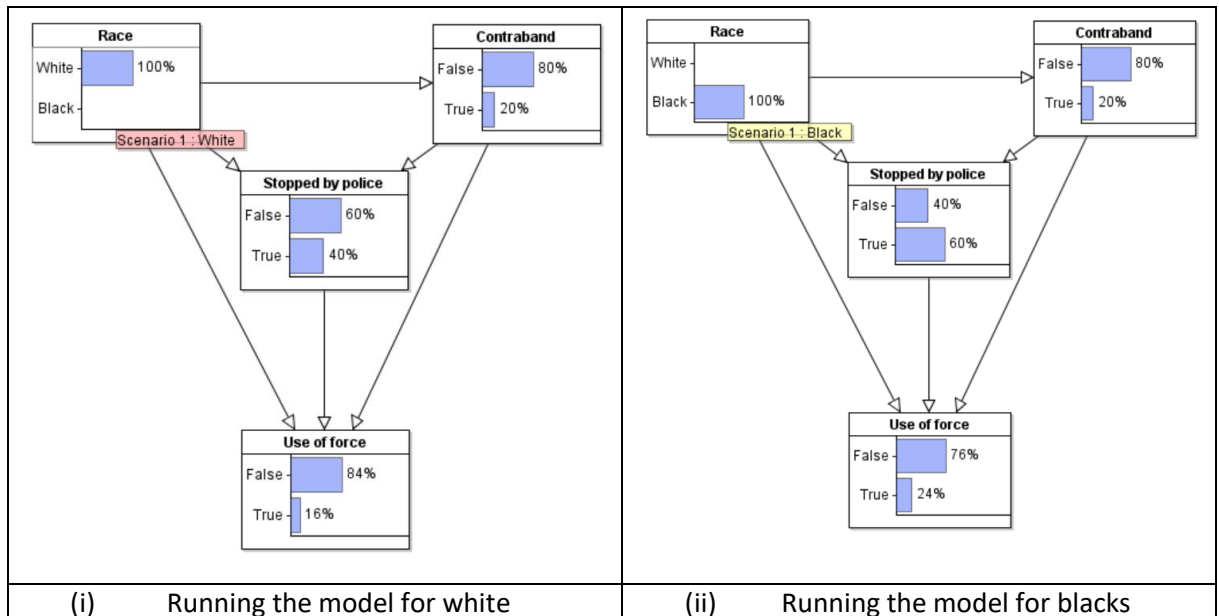

| (i) Running the model for white | (ii) Running the model for blacks |

*Figure 3 Model confirms there is bias*

Note that for whites, there is a 16% probability of use of force, compared to a 24% for blacks.

However, we can use the model to show what happens when the dataset is restricted to considering only cases where people are stopped by police i.e. where our dataset is censored. To encode this, we simply enter 'True' for the node "Stopped by police" as shown in Figure 4 and this effectively makes all variable outcomes in the model dependant on this 'constraint'.

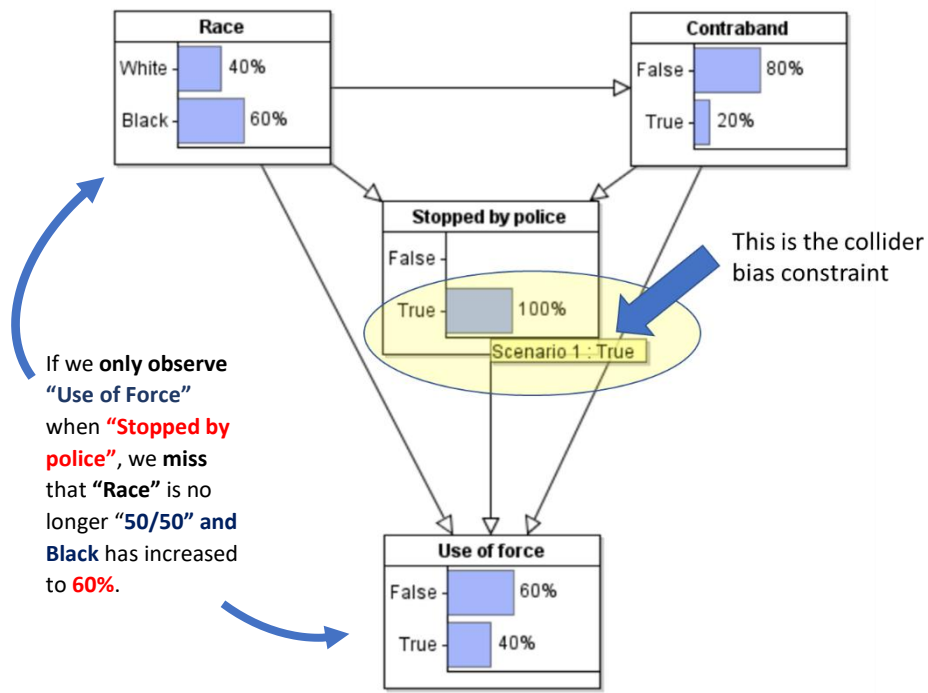

If we **only observe** **"Use of Force"** when **"Stopped by police"**, we **miss** that **"Race"** is no longer "**50/50" and Black** has increased to **60%**.

*Figure 4 Collider bias introduced when we restrict the study to people stopped by police*

Note that when we introduce this 'constraint' the dataset no longer has an equal number of blacks as whites. With this constrained dataset, there no longer appears to be any bias – as shown by running the model for whites and blacks respectively in Figure 5.

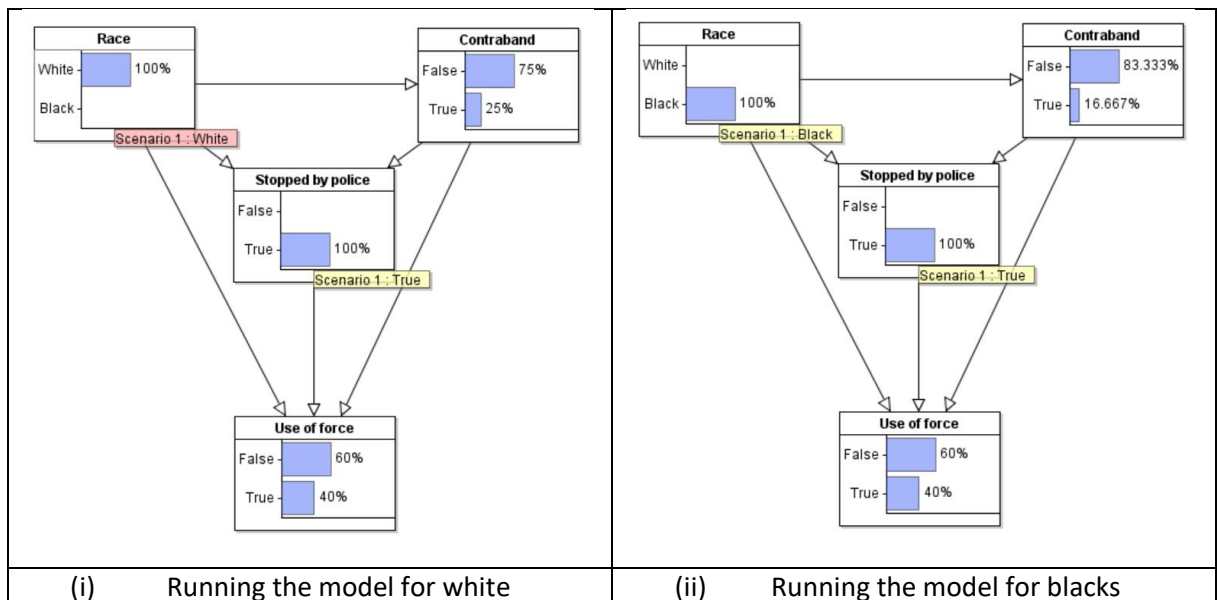

| (i) Running the model for white | (ii) Running the model for blacks |

*Figure 5 Model confirms there is 'no bias' for the restricted dataset*

For both whites and blacks the probability of force being used is the same: 40%. However, note that the model also provides an explanation for the result. In the restricted dataset whites are more likely to be carrying contraband (25%) than blacks (16.67%). This is because 6/10 stops are for blacks and 4/10 are for whites. Additionally, what constitutes the likelihood of "contraband" for whites is lower than that for blacks. This excess stoppage of blacks may not be efficient in "capturing" criminals and potentially "antagonistically criminalizes" innocent people.

As (Knox, Lowe, and Mummolo 2020) argue, the example shows both the danger of restricting a study to data which itself may be inherently biased (in this case the data of people stopped by police) but also how to avoid the problem by simply adding the necessary missing information and running the inference (as we do here in a BN which we believe is simpler than described by Knox et al). In this case the missing information are the numbers required for the conditional probability table of the node 'Stopped by police'. There is no reason why these numbers (or good estimates of them) should not be available.

### 3. Using causal models to consider alternative hypotheses

One great strength of the Bayesian causal modelling approach is its simple graphical "synthesis" of a complex problem via a plurality of competing models that can be entertained as explanations for the observed data. Each of these models may contain different sets of assumptions that might hypothetically explain the mechanism by which the data are generated. Beliefs in and about the extent and importance of these different mechanisms drives the ideological, criminological, and legal debates surrounding this issue. We might be able to entertain several possible relevant mechanisms at play, including:

- Consider anti-social misbehaviour. It might be argued that misbehaviour attracts police attention independently of whether contraband is present or not. If different races (or age groups) tend to visibly misbehave at different rates, then this will increase the likelihood of being stopped by the police independently of whether they have contraband.
- If we disregard tip-offs or intelligence led policing, then whether contraband is discovered depends on whether misbehaviour has taken place and not solely on whether contraband is present.
- Police defence in court is often reliant on establishing that the victim of a use of force was misbehaving into some way (perhaps threateningly), simply because in establishing a motive for the police stop it provides a legal defence against competing reasons for the stop (such as racism). Similar logic applies if contraband is present.
- Rather than assume racism as an implicit hypothesis, whether police are racist or not might be explicitly modelled in the BN as a node itself. Given this a proxy for racist police might be used, such on whether the police run patrols with mix race officers etc.

When we combine some or all of these, we end up with a plethora of possible models that might explain the mechanism at play here. Two possible mechanisms are shown in Figure 6, but it is easy to imagine many more.

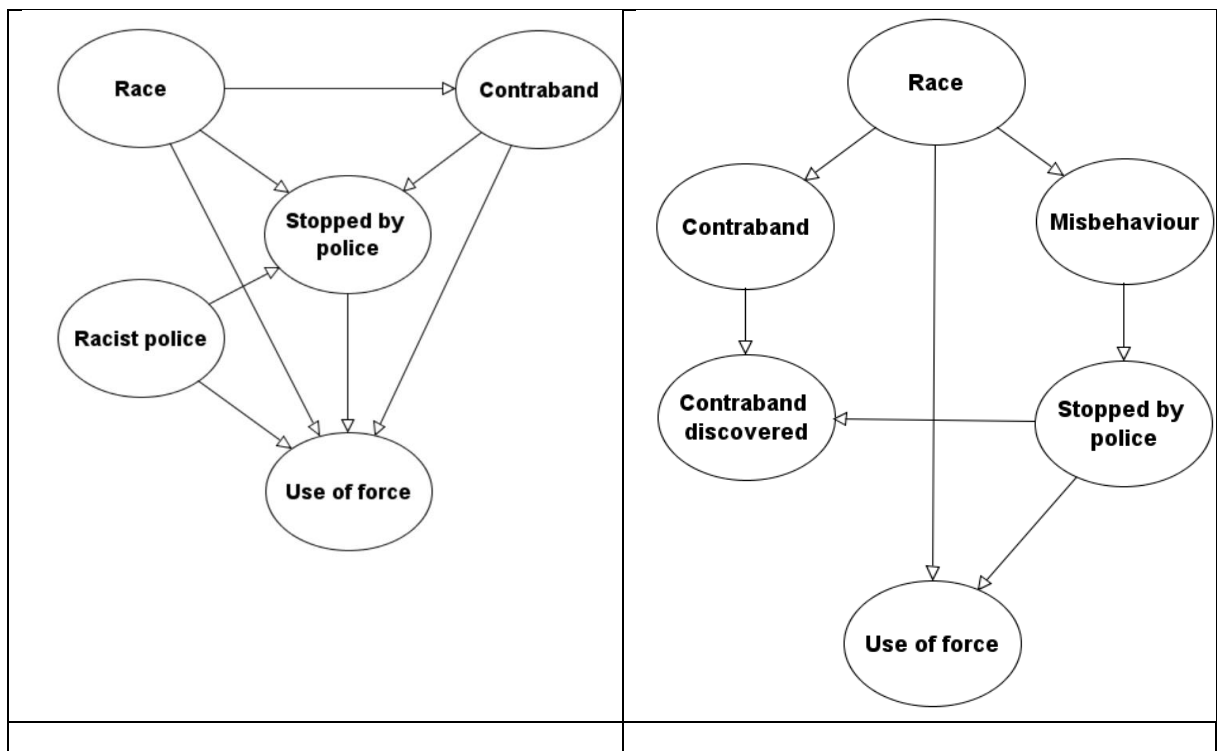

*Figure 6 Two different models with different causal mechanisms*

## 4. Conclusion

We have shown that the live social arguments surrounding the use of force and potentially racist intent, whether they have roots in scientific, ideological, or political differences, can be posed in the language of cause and effect. Doing so clarifies the differences in the arguments made and identifies what variables, and how they relate to other variables, might be defined, and measured. Once this is done our social arguments can be made testable by empirical means. And again, Bayesian modelling can provide the rational tools to carry out this process, as demonstrated in our recent paper showing how competing legal arguments presented in court can be represented and adjudged (Neil et al. 2019).

**References**

Agena Ltd. 2020. "AgenaRisk." http://www.agenarisk.com.

Bickel, P J, E A Hammel, and J W O'Connell. 1975. "Sex Bias in Graduate Admissions: Data from Berkley." *Science* 187: 398–404.

Bronner, L. 2020. "Why Statistics Don't Capture The Full Extent Of The Systemic Bias In Policing." *FiveThirtyEight*, June 25, 2020. https://fivethirtyeight.com/features/why-statistics-dont-capture-the-full-extent-of-the-systemic-bias-in-policing/.

Fenton, Norman E., Martin Neil, and Anthony Constantinou. 2019. "Simpson's Paradox and the Implications for Medical Trials," December. http://arxiv.org/abs/1912.01422.

Fenton, Norman E. 2020. "A Note on 'Collider Bias Undermines Our Understanding of COVID-19 Disease Risk and Severity' and How Causal Bayesian Networks Both Expose and Resolve the Problem," May. http://arxiv.org/abs/2005.08608.

Fenton, Norman E, and Martin Neil. 2018. *Risk Assessment and Decision Analysis with Bayesian Networks*. 2nd ed. CRC Press, Boca Raton.

Fryer, Roland G. 2016. "An Empirical Analysis of Racial Differences in Police Use of Force." Cambridge, MA. https://doi.org/10.3386/w22399.

———. 2019. "An Empirical Analysis of Racial Differences in Police Use of Force." *Journal of Political Economy* 127 (3): 1210–61. https://doi.org/10.1086/701423.

Knox, Dean, Will Lowe, and Jonathan Mummolo. 2020. "Administrative Records Mask Racially Biased Policing." *American Political Science Review*, May, 1–19. https://doi.org/10.1017/S0003055420000039.

Neil, Martin, Norman E Fenton, David A Lagnado, and Richard Gill. 2019. "Modelling Competing Legal Arguments Using Bayesian Model Comparison and Averaging." *AI and Law* 27: 403–30. https://doi.org/10.1007/s10506-019-09250-3.

Patty, B, and J Hanson. 2020. "Race as a Factor in Police Homicides." https://securitystudies.org/wp-content/uploads/2020/06/Race-as-a-factor-in-Police-Homicides-062320.pdf.

Pearl, Judea. 2000. *Causality: Models Reasoning and Inference*. Cambridge University Press.

Pearl, Judea, and Dana Mackenzie. 2018. *The Book of Why : The New Science of Cause and Effect*. New York: Basic Books.

Ross, Cody T. 2015. "A Multi-Level Bayesian Analysis of Racial Bias in Police Shootings at the County-Level in the United States, 2011–2014." Edited by Peter James Hills. *PLOS ONE* 10 (11): e0141854. https://doi.org/10.1371/journal.pone.0141854.

Ross, Cody T., Bruce Winterhalder, and Richard McElreath. 2018. "Resolution of Apparent Paradoxes in the Race-Specific Frequency of Use-of-Force by Police." *Palgrave Communications* 4 (1): 61. https://doi.org/10.1057/s41599-018-0110-z.

Simpson, E. H. 1951. "The Interpretation of Interaction in Contingency Tables." *Journal of the Royal Statistical Society B* 2: 238–41.